\documentclass[twocolumn]{aastex62}
\usepackage{lineno}
\usepackage{natbib}
\usepackage{amsmath}
\usepackage{subfigure}
\setcitestyle{sort,comma}
\usepackage{csvsimple}

\usepackage[none]{hyphenat}

\graphicspath{{./}{figures/}}

\shorttitle{Low dayside emission on TOI-431\,b}

\begin{document}

\title{\large{Low 4.5$\,\mu$m  Dayside Emission Disfavors a Dark Bare-Rock scenario\\ for the Hot Super-Earth TOI-431\,b}}

\correspondingauthor{Christopher Monaghan}
\email{christopher.monaghan@umontreal.ca}

\author[0009-0005-9152-9480]{Christopher Monaghan} 
\affil{Department of Physics and Trottier Institute for Research on Exoplanets, Universit\'{e} de Montr\'{e}al, Montreal, QC, Canada}

\author[0000-0001-6809-3520]{Pierre-Alexis Roy} 
\affil{Department of Physics and Trottier Institute for Research on Exoplanets, Universit\'{e} de Montr\'{e}al, Montreal, QC, Canada}

\author[0000-0001-5578-1498]{Bj\"{o}rn Benneke}
\affil{Department of Physics and Trottier Institute for Research on Exoplanets, Universit\'{e} de Montr\'{e}al, Montreal, QC, Canada}
\affil{Department of Earth, Planetary, and Space Sciences, University of California, Los Angeles, Los Angeles, CA, USA}

\author{Ian J.\ M.\ Crossfield}
\affiliation{Department of Physics and Astronomy, University of Kansas, Lawrence, KS, USA}
\affiliation{Max Planck Institut f\"ur Astronomie, K\"onigstuhl 17, 69117, Heidelberg, Germany}

\author[0000-0002-2195-735X]{Louis-Philippe Coulombe} 
\affil{Department of Physics and Trottier Institute for Research on Exoplanets, Universit\'{e} de Montr\'{e}al, Montreal, QC, Canada}

\author[0000-0002-2875-917X]{Caroline Piaulet-Ghorayeb}
\altaffiliation{E. Margaret Burbridge Postdoctoral Fellow}
\affiliation{Department of Astronomy \& Astrophysics, University of Chicago, 5640 South Ellis Avenue, Chicago, IL 60637, USA}
\affil{Department of Physics and Trottier Institute for Research on Exoplanets, Universit\'{e} de Montr\'{e}al, Montreal, QC, Canada}

\author[0000-0003-0514-1147]{Laura Kreidberg}
\affil{Max Planck Institut f\"ur Astronomie, K\"onigstuhl 17, 69117, Heidelberg, Germany}

\author[0000-0001-8189-0233]{Courtney D. Dressing}
\affiliation{Department of Astronomy, University of California, Berkeley, Berkeley, CA 94720}

\author[0000-0002-7084-0529]{Stephen R. Kane}
\affiliation{Department of Earth and Planetary Sciences, University of California, Riverside, CA 92521, USA}

\author[0000-0003-2313-467X]{Diana Dragomir}
\affil{Department of Physics and Astronomy, University of New Mexico, Albuquerque, NM, USA}

\author[0000-0003-4990-189X]{Michael W. Werner}
\affil{Jet Propulsion Laboratory, California Institute of Technology, 4800 Oak Grove Drive, Pasadena, CA 91109, USA}

\author[0000-0001-9521-6258]{Vivien Parmentier}
\affil{Atmospheric, Oceanic \& Planetary Physics, Department of Physics, University of Oxford, Oxford OX1 3PU, UK}

\author[0000-0002-8035-4778]{Jessie L. Christiansen}
\affil{Caltech/IPAC-NExScI, M/S 100-22, 1200 E. California Blvd, Pasadena, CA 91125, USA}

\author[0000-0001-9414-3851]{Farisa Y. Morales}
\affiliation{Jet Propulsion Laboratory, California Institute of Technology, 4800 Oak Grove Drive, Pasadena, CA 91109, USA}

\author[0000-0001-6298-412X]{David Berardo}
\affiliation{Department of Earth, Atmospheric and Planetary Sciences, Massachusetts Institute of Technology, Cambridge, MA 02139, USA}
\affiliation{Department of Physics and Kavli Institute for Astrophysics and Space Research, Massachusetts Institute of Technology, Cambridge, MA 02139, USA}

\author[0000-0002-8990-2101]{Varoujan Gorjian}
\affiliation{Jet Propulsion Laboratory, California Institute of Technology, 4800 Oak Grove Drive, Pasadena, CA 91109, USA}

\begin{abstract}

The full range of conditions under which rocky planets can host atmospheres remains poorly understood, especially in the regime of close-in orbits around late-type stars. One way to assess the presence of atmospheres on rocky exoplanets is to measure their dayside emission as they are eclipsed by their host stars. Here, we present Spitzer observations of the 4.5\,µm secondary eclipses of the rocky super-Earth TOI-431\,b, whose mass and radius indicate an Earth-like bulk composition (3.07\,$\pm 0.35$\,M$_{\oplus}$, 1.28\,$\pm 0.04$\,R$_{\oplus}$). Exposed to more than 2000 times the irradiation of Earth, dayside temperatures of up to 2400~K are expected if the planet is a dark bare-rock without a significant atmosphere. Intriguingly, despite the strong stellar insolation, we measure a secondary eclipse depth of only 33$\pm$22\,ppm, which corresponds to a dayside brightness temperature of $1520_{-390}^{+360}$\,K. This notably low eclipse depth disagrees with the dark bare-rock scenario at the 2.5\,$\sigma$ level, and suggests either that the planet is surrounded by an atmosphere, or that it is a bare-rock with a highly reflective surface. In the atmosphere scenario, the low dayside emission implies the efficient redistribution of heat to the nightside, or by molecular absorption in the 4--5\,$\mu$m bandpass. In the bare-rock scenario, a surface composition made of a high-albedo mineral species such as ultramafic rock can lead to reduced thermal emission consistent with low eclipse depth measurement. Follow-up spectroscopic observations with the James Webb Space Telescope hold the key to constraining the nature of the planet.

\end{abstract}

\keywords{Exoplanets (498); Exoplanet atmospheres (487); Planetary atmospheres (1244)}

%% From the front matter, we move on to the body of the paper.
%% Sections are demarcated by \section and \subsection, respectively.
%% Observe the use of the LaTeX \label
%% command after the \subsection to give a symbolic KEY to the
%% subsection for cross-referencing in a \ref command.
%% You can use LaTeX's \ref and \label commands to keep track of
%% cross-references to sections, equations, tables, and figures.
%% That way, if you change the order of any elements, LaTeX will
%% automatically renumber them.
%%
%% We recommend that authors also use the natbib \citep
%% and \citet commands to identify citations.  The citations are
%% tied to the reference list via symbolic KEYs. The KEY corresponds
%% to the KEY in the \bibitem in the reference list below. 
% \tableofcontents

\section{Introduction} \label{sec:intro}

%Recent advancements in exoplanetary science have allowed for further investigation into the nature of rocky exoplanet atmospheres. Notably, 
The ability for rocky exoplanets to retain sizable atmospheres under different conditions remains a key area of research. Although rocky planets are unlikely to retain their primordial hydrogen envelope, they may generate secondary atmospheres at later stages in their life \citep{Kite_2020, tian2024atmosphericchemistrysecondaryhybrid}. Determining which rocky exoplanets host atmospheres and studying their properties will allow us to investigate how these atmospheric phenomena impact planetary evolution. 

A number of processes are known to affect the atmospheric retention of exoplanets. Mass loss is driven by many factors, including photoevaporation and stellar wind erosion caused by high levels of irradiation from the host star \citep[e.g.,][]{kubyshkina2024planetaryatmospherestimeeffects}. Core-powered mass loss can also play a role after formation \citep{Ginzburg_2016, Ginzburg_2018}. The age of the planet is also a key factor in determining an atmospheric presence, as older planets ($>$ 6 Gyr) may have ceased degassing processes, halting atmospheric replenishment \citep[e.g.,][]{2022ApJ...930L...6U}. 

The compositions of secondary atmospheres may impact the rate of atmospheric mass loss. For example, the initial content of volatiles in the star forming region and associated protoplanetary disk will allow for the capture of gaseous species during the planet's evolution \citep{Wordsworth_Kreidberg_2022}. Furthermore, volcanism is known to affect the composition of planetary atmospheres through the outgassing of interior volatiles \citep{Liggins_Jordan_Rimmer_Shorttle_2022, Liggins_Jordan_Rimmer_Shorttle_2023}. Studying the ways in which these processes interact is critical to understanding the conditions for atmospheric retention.

Detecting the atmospheres of warm exoplanets is readily achieved using thermal emission spectroscopy, which provides the most promising method to constrain the surface composition of rocky planets \citep{2018haex.bookE..40A, Whittaker_2022}. Such studies require distinguishing the contribution of the exoplanet's light from the combined system during the planet's secondary eclipse. Observing the occultations and phase curves of rocky planets in the infrared regime allows us to detect the presence of an atmosphere by measuring the brightness temperature of the planet \citep{Koll_2019}. These observations are especially favorable for hot exoplanets orbiting late-type stars as the higher dayside temperatures result in larger eclipse depths.

\subsection{Thermal emission spectroscopy of terrestrial bodies}

%A surprisingly high number of exoplanets have been found with short orbital periods \textbf{(Dawson \& Johnson 2018, Fulton \& Petigura 2018)}. More than 100 detected exoplanets fall into the extreme subclass of ultra short period (USP) planets, with orbital periods less than one day \textbf{(Winn et al. 2018)}. The majority of USPs are found to be sub-Earths or super-Earths of $R < 2R_{\oplus}$ \textbf{(e.g. Sanchis-Ojeda et al. 2014, Uszoy et al. 2021)}. The exact formation pathways of rocky USPs are still under study, although it is possible that such planets are formed by the photoevaporation of closely orbiting sub-Neptunes \textbf{(Xiu-ming \& Jiang-hui 2020)}.

Discovering which rocky exoplanets have atmospheres will further our understanding of the factors causing atmosphere loss and retention on terrestrial bodies. Close-in terrestrial planets are more susceptible to atmospheric loss due to the high irradiation received from their host stars, and are unlikely to host a significant gaseous envelope \citep{Lopez_2017}. However, highly irradiated hot planets may host molten surfaces, which could maintain volatile-rich atmospheres from the outgassing of lava \citep[e.g.,][]{Hu_Bello-Arufe_Zhang_Paragas_Zilinskas_Van_Buchem_Bess_Patel_Ito_Damiano_et_al._2024, Patel_Brandeker}. Particularly hot planets may instead host silicate atmospheres through the evaporation of surface magma \citep[e.g.,][]{Schaefer_Fegley_2009, Maurice_Dasgupta_Hassanzadeh_2024}.

The present-day catalog of eclipse observations of rocky planets has produced a diverse range of results, and enforces the need for further research into how terrestrial planets develop and retain secondary atmospheres. Many planets observed fall into the subclass of Ultra Short Period (USP) planets, with orbital periods shorter than one day \citep{Sanchis_Ojeda_2014, Winn_Sanchis-Ojeda_Rappaport_2018}. One of the most well-studied USPs, 55 Cnc e, has been central to a number of emission-based studies in the infrared regime \citep[e.g.,][]{Demory_2015, Tsiaras_2016, rasmussen2023nondetectionironhighresolutionemission, Patel_Brandeker}. Recent observations by the James Webb Space Telescope (\textit{JWST}) suggest that 55 Cnc e has a volatile rich atmosphere comprised mostly of CO$_{2}$ or CO fueled by the outgassing of a magma ocean \citep{Hu_Bello-Arufe_Zhang_Paragas_Zilinskas_Van_Buchem_Bess_Patel_Ito_Damiano_et_al._2024,Patel_Brandeker}. K2 and Spitzer phase curve observations of another USP planet, K2-141 b, have previously suggested the presence of a tenuous rock vapor atmosphere caused by a molten dayside \citep{Zieba_Zilinskas_Kreidberg_Nguyen_Miguel_Cowan_Pierrehumbert_Carone_Dang_Hammond_et_al._2022}. 
In contrast, Spitzer observations of the USP planets LHS 3844 b and GJ 1252 b have shown these planets to have thermal emissions that are largely consistent with bare-rocks without an atmosphere \citep{Kreidberg_Koll_Morley_Hu_Schaefer_Deming_Stevenson_Dittmann_Vanderburg_Berardo_et_al._2019,Kane2020,Crossfield_Malik_Hill_Kane_Foley_Polanski_Coria_Brande_Zhang_Wienke_et_al._2022}.
Similarly, \textit{JWST} observations of a number of earth-sized planets have found a lack of significant evidence for an atmospheric presence over a large range of dayside temperatures \citep{Greene_Bell_Ducrot_Dyrek_Lagage_Fortney_2023, Zieba_Kreidberg, Xue_Bean_Zhang_Mahajan_Ih_Eastman_Lunine_Mansfield_Coy_Kempton_et_al._2024, Zhang_Hu_Inglis_Dai_Bean_Knutson_Lam_Goffo_Gandolfi_2024,Mansfield_Xue_Zhang_Mahajan_Ih_Koll_Bean_Coy_Eastman_Kempton_et_al._2024, wachiraphan2024thermalemissionspectrumnearby, luque2024darkbarerocktoi1685, Ducrot_2024}. Recent studies of the super-Earth LHS 1478 b, however, observed a notably low eclipse depth inconsistent with a bare-rock model \citep{august2024hotrockssurveyi}. Thus, while the majority of terrestrial exoplanets observed so far are consistent with bare-rocks, the reasons why some may be able to retain a secondary atmosphere remains largely unknown. 

%Of the 10 planets with direct infrared flux measurements, only 55 Cnc e and TOI-431\,b show evidence of atmospheric heat redistribution. Recent observations by the JWST suggest that 55 Cnc e has a volatile rich atmosphere comprised mostly of CO$_{2}$ or CO fueled by the outgassing of a magma ocean \textbf{(Hu et al. 2024, Patel et al. 2024)}. The remaining 8 planets have spectra largely consistent with bare-rock models or tenuous, thin atmospheres with limited heat recirculation. Initial reports by \textbf{Zieba et al. (2022)} suggested a tenuous rock vapor atmosphere to be present on K2-141b, although more recent observations of its phase curve are largely consistent with a bare-rock \textbf{IN PREP}.

\subsection{TOI-431\,b}

\begin{table}
\centering
\footnotesize
\begin{tabular}{ l   ll c}
\hline
\hline
Parameter & Unit & Value & Ref \\
\hline
Stellar Parameters \\
\\
Stellar Mass M$_{*}$ & M$_{\odot}$ & 0.78 $\pm$ 0.07 & 1 \\ 
Stellar Radius R$_{*}$ & R$_{\odot}$ & 0.731 $\pm$ 0.022 & 1 \\ 
Effective Temp. T\tiny{eff} & K & 4850 $\pm$ 75 & 1 \\
\\
\hline
Planetary Parameters \\
\\
Mass M$_{p}$ & M$_{\oplus}$ & $3.07\pm0.35$ & 1 \\
Radius R$_{p}$ & R$_{\oplus}$ & $1.28\pm0.04$ & 1 \\
Bulk density $\rho$ & g/cm$^{3}$ & $8.0\pm1.0$ & 1 \\
Equilibrium Temp. T\tiny{eq} & K & $1862\pm42$ & 1 \\
Max dayside Temp. T\tiny{d} & K & $2400\pm60$ & 2 \\
Period P & d & $0.490\:047_{-0.000\: 007}^{+0.000\:010}$ & 1\\
Semi-major Axis a & AU & $0.0113_{-0.0003}^{+0.0002}$ & 1\\
Impact Parameter b & & $0.34_{-0.06}^{+0.07}$ & 1\\
Scaled semi-major axis & & $3.324\pm0.13$ & 2 \\
Transit time t$_{0}$ & BJD$_{TDB}$ & $2458627.538_{-0.002}^{+0.003}$ & 1 \\
Secondary Eclipse t\tiny{sec} & BJD$_{TDB}$ & $2458627.783_{-0.002}^{+0.003}$ & 2 \\
\\
\hline
\hline
\end{tabular}
\caption{\label{tab:params} Summary of stellar and planetary parameters of TOI-431\,b, with error bars. \newline 1: \citet{Osborn_2021} \newline 2: This work, derived}
\end{table}

The USP super-Earth TOI-431\,b is a prime target for infrared thermal emission measurements. The planet has a mass of 3.07 $\pm$ 0.35 M$_{\oplus}$, a radius of 1.28 $\pm$ 0.04 R$_{\oplus}$, and a period of 0.49 days \citep[Table \ref{tab:params}, ][]{Osborn_2021}. TOI-431\,b has an Emission Spectroscopy Metric \citep[ESM, ][]{Kempton_Bean_Louie_Deming_Koll_Mansfield_Christiansen} of approximately 16 and orbits a cool K-type star, suggesting the potential for significant flux ratios to be measured during an occultation event \citep{hord2023identificationtessobjectsatmospheric}. 

TOI-431 b is expected to lack an atmospheric presence, as XUV irradiation 
from the host star should drive an incredibly high mass-loss rate due to the close-in orbit \citep{King_Corrales_fern_Wheatley_Malsky_Osborn_Armstrong_2024, jiang2025estimatemassescapingrates}. Assuming a molten surface is present, recent simulations of TOI-431\,b's emission spectra suggest an eclipse depth between 95-120ppm to be measured at 4.5\,µm, depending on the planet's surface temperature \citep{Seidler_Sossi_Grimm_2024}. A similarly high eclipse depth is expected if the planet is consistent with a dark bare-rock. 

TOI-431\,b was discovered orbiting a K type star alongside the sub-Neptune TOI-431 d in 2021 by \citet{Osborn_2021} using transit observations from the Transiting Exoplanet Survey Satellite \citep[TESS,][]{Ricker_2014}. A third non-transiting planet, TOI-431 c, was also discovered by \citet{Osborn_2021} using high precision Doppler spectroscopy from the High Accuracy Radial velocity Planet Searcher \citep[HARPS,][]{2002Msngr.110....9P} and the HIgh REsolution Spectrograph \citep[HIRES,][]{1994SPIE.2198..362V}. The system may serve as a key target in future studies of planetary evolution due to its unique orbital configuration, having one non-transiting planet between two transiting planets. Furthermore, TOI-431\,b and TOI-431 d orbit on opposing sides of the radius-period valley, providing a rare opportunity to study the dichotomy between the evolution of planets in the same system on opposite ends of the gap \citep[e.g.,][]{2017AJ....154..109F}. 

Here, we present Spitzer eclipse observations of the USP super-Earth TOI-431\,b, combined with an analysis of the planet's potential atmosphere. In Section \ref{sec:obs_red}, we describe our analysis of the Spitzer observations and measure the eclipse depth and dayside brightness temperature of TOI-431\,b. Section \ref{sec:atms} discusses these measurements and compares the results to both bare-rock and atmospheric models of different compositions. The significance of our occultation depth and the results of our modeling are further discussed in Section \ref{sec:dis}. Finally, in Section \ref{sec:context}, we present an analysis of TOI-431\,b in the context of other thermal emission observations of rocky exoplanets, and consider future prospects of emission spectroscopy.

\begin{table}[t!]
\centering
\footnotesize
\begin{tabular}{c c c c}
\hline
AOR    & Date &    Eclipse Depth & Error $\pm\sigma$\\
 & YYYY-MM-DD & (ppm) &  (ppm)\\
\hline
\hline
69969664 & 2019-07-09 & 59 & 57\\
69969408 & 2019-07-19 & 105 & 59\\
69969152 & 2020-01-03 & 37 & 59\\
69968896 & 2020-01-06 & -71 & 63\\
69968640 & 2020-01-07 & 51 & 57\\
69968128 & 2020-01-08 & 8 & 56\\
69967104 & 2020-01-09 & 28 & 56\\
\hline
\multicolumn{2}{l}{\textbf{Weighted Average}} & \textbf{33} & \textbf{22} \\
 
\hline
\hline
\end{tabular}
\caption{\label{tab:obs} Summary of Spitzer/IRAC 4.5\,µm visits and measured eclipse depths and $\sigma$ error bars as calculated from the Pixel-level Decorrelation (PLD) reduction in ExoTEP.}
\end{table}

\section{Observations and data reduction}\label{sec:obs_red}

\subsection{Eclipse Observations}
A total of 7 eclipses of TOI-431\,b were observed at 4.5\,µm using the Spitzer Space Telescope Infrared Array Camera (IRAC) in Channel 2. These observations were taken as part of the large TESS planets follow-up program \citep[GO 14084,][]{2018sptz.prop14084C}. We targeted this planet specifically because it was one of the highest-ESM rocky planets known while Spitzer was still operating. Two of these events were observed in July 2019, while the remaining five were taken in January 2020, shortly before Spitzer was decommissioned. Each observation was a continuous and near-identical exposure of $\sim$5.24h centered approximately on the center of the predicted time of the secondary eclipse, allowing for $\sim$2 hours of both pre-eclipse and post-eclipse baseline.

The Spitzer/IRAC images were processed using standard methods outlined in \citet{2017ApJ...834..187B} and \citet{2019ApJ...887L..14B}. We start from the flat-fielded and dark-subtracted ``Basic Calibrated Data" images, before using the method outlined in \citet{2015ApJ...810..118K} for background subtraction. The position of the star is determined using flux-weighted centroiding with a radius of 3.5 pixels \citep{2019NatAs...3..813B}. The aperture, trim duration, and bin size are then chosen to minimize the RMS of the unbinned residuals. The seven observations and their results from the following data analysis are summarized in Table \ref{tab:obs}.

\begin{figure}[t!]
\centering
\includegraphics[width=\linewidth]{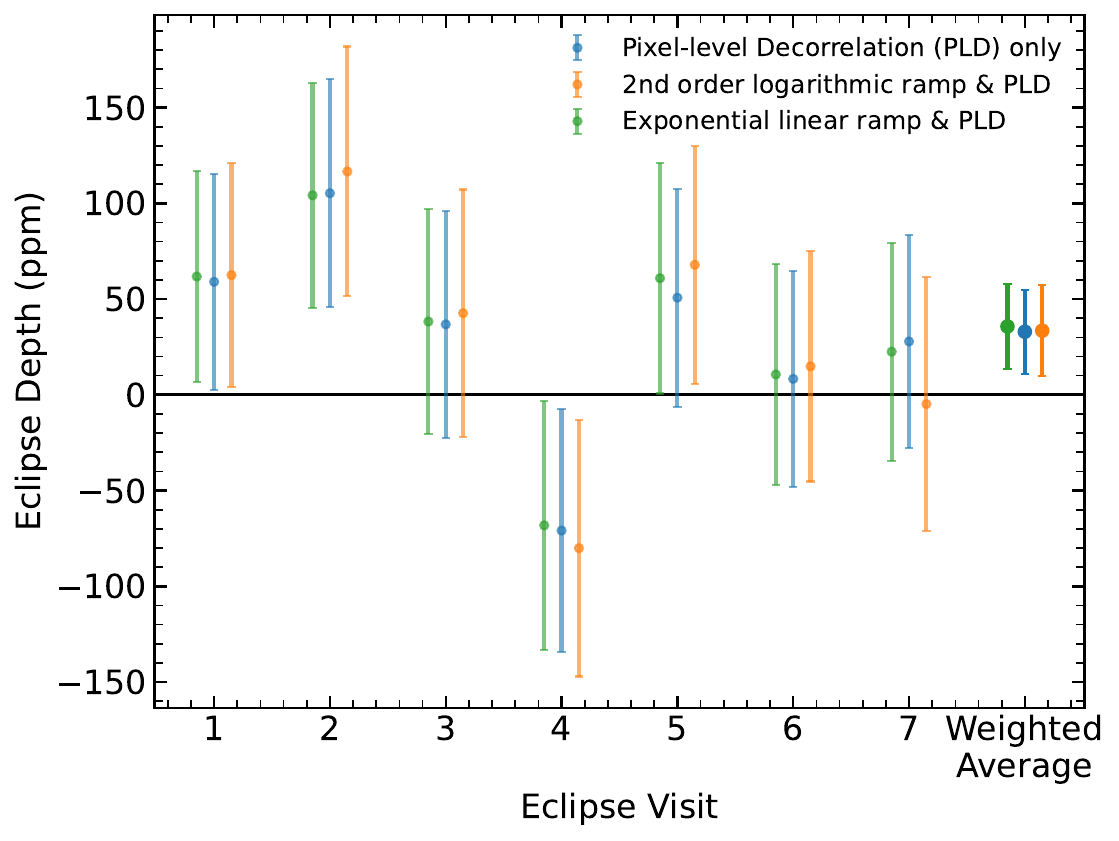}
\caption{Measured eclipse depths for all seven visits using three different systematic models, including the associated error bars of each. The rightmost points show the weighted average across all visits for the individual systematics.}
\label{fig:systematicscomparison}
\end{figure}

\subsection{Data Analysis}
We analyzed the Spitzer light curves using the modular Exoplanet Transits Eclipses \& Phasecurves (ExoTEP) framework \citep[e.g.,][]{2017ApJ...834..187B, 2019ApJ...887L..14B, 2019NatAs...3..813B, benneke2024jwstrevealsch4co2, Roy_Benneke_Piaulet_Crossfield_Kreidberg_Dragomir_Deming_Werner_Parmentier_Christiansen_et_al._2022, Roy_2023, Piaulet_Ghorayeb_2024}. ExoTEP employs a Markov Chain Monte Carlo (MCMC) algorithm to fit the secondary eclipse light curve and both astrophysical and systematic parameters to the observed data. The emcee python package was used to explore the joint posterior distributions of the parameters from uniform priors using the MCMC \citep{2013PASP..125..306F}. For each visit, we perform an individual MCMC with 50,000 steps to fit the eclipse depth, photometric scatter, and systematic parameters. The remaining astrophysical parameters were set to values from \citet{Osborn_2021}. Given that the eccentricity of TOI-431\,b is consistent with 0 \citep{Osborn_2021}, we do not fit for the time of the secondary eclipse, and instead use a value of $T_{sec} = T_{t} + \frac{1}{2}P$ for all 7 visits.

Considering TOI-431\,b’s short orbit and the 5 hour duration of each visit, we included a phase curve component in our light-curve modeling to account for the changing visibility of the dayside during the planet’s egress and ingress, assuming negligible nightside emissions. Following \citet{Parviainen_2022}, the lightcurve was modeled as a sine wave between $1$ and $1 + F_{p}$ to account for the phase of the planet's dayside during the visits:

\begin{figure}[t]
\centering
\includegraphics[width=\linewidth]{{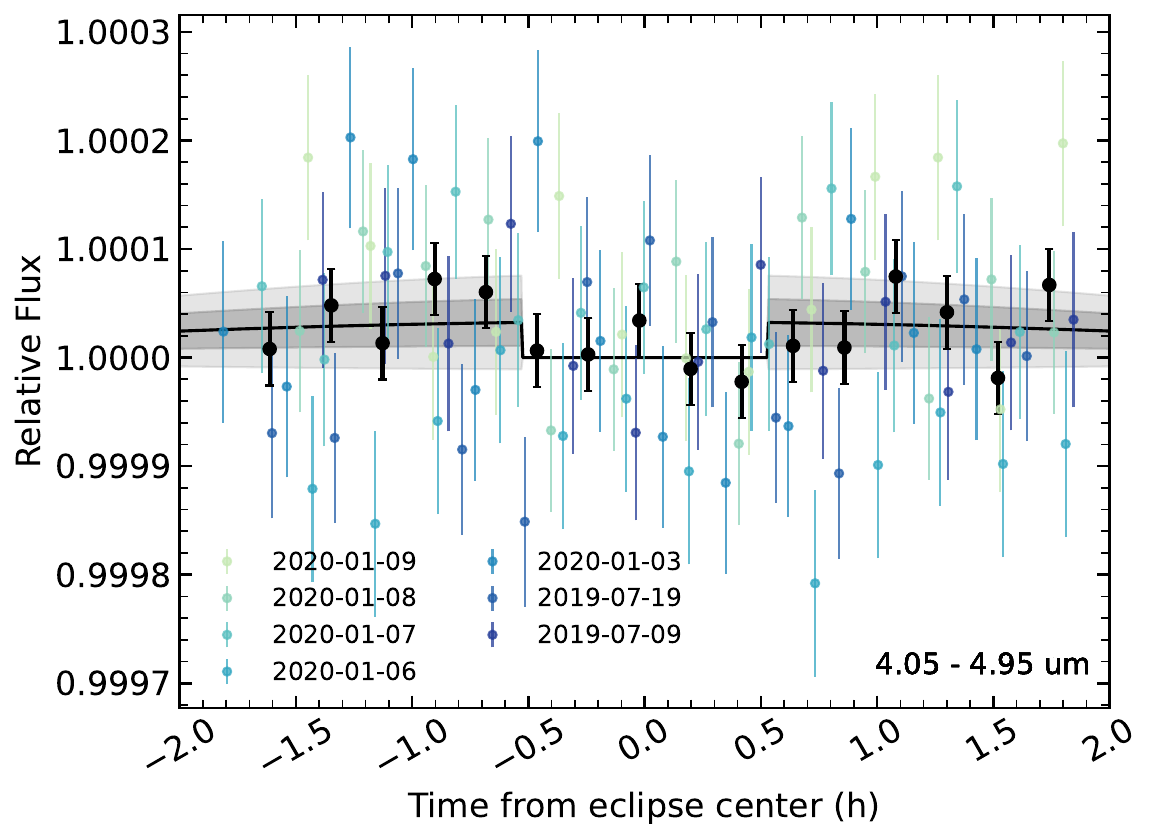}}
\caption{Best-fitting white light curve from the analysis of the seven eclipses of TOI-431\,b with Spitzer Ch 2. The eclipse light curve generated from the average eclipse depth of 33ppm is shown in the black line, with $\pm1\sigma$ and $\pm2\sigma$ ranges shown in gray. The blue circles represent the individual systematics-corrected observations and their associated error bars. The Spitzer data points are binned to 8.33 minute intervals, and the bold black points represent the combination of all the data binned by groups of 250 points. The first and last 45 minutes of each data set have been removed.}
\label{fig:bestfit}
\end{figure}

\begin{equation}
    F = \frac{F_{p}}{2}\mathrm{cos}\left(\frac{2\pi t}{P}+\pi\right) + \left(1 + \frac{F_{p}}{2}\right)
\end{equation}

\noindent where $F$ represents the relative flux of the system, $F_{p}$ the eclipse depth of the visit, $P$ the orbital period, and $t$ the time. Similarly sinusoidal lightcurves have been modeled for other super-Earths \citep[e.g.,][]{Rouan_2011, Kreidberg_Koll_Morley_Hu_Schaefer_Deming_Stevenson_Dittmann_Vanderburg_Berardo_et_al._2019, Zieba_Zilinskas_Kreidberg_Nguyen_Miguel_Cowan_Pierrehumbert_Carone_Dang_Hammond_et_al._2022}.

We used a Pixel Level Decorrelation (PLD) model to account for the systematics associated with sensitivity variation and stellar motion on the detector \citep{Deming_Knutson_Kammer_Fulton_Ingalls_Carey_Burrows_Fortney_Todorov_Agol_et_al._2015}. The analysis was performed by binning the data into 20s intervals. Two systematic models were implemented on top of the PLD to account for the effects of drift and test for variation: one with a second-order logarithmic ramp in time, and one with an exponential-linear ramp in time \citep[e.g.,][]{May_2022}. However, by cutting off the first 45 minutes of each data set, we found that the systematic drift was effectively removed, and subsequent analyses showed that the three systematic models produced results consistent with one another (Figure \ref{fig:systematicscomparison}). As such, we used a model with no systematic ramp to reduce the overall complexity of the model and avoid over-fitting for a limited drift. 

Using the methodology outlined above, we perform our ExoTEP analysis on all 7 visits fitting for the eclipse depth, photometric scatter, and PLD coefficients in each visit. From our MCMC analysis of the light curve, we calculate the weighted average across all 7 visits and measure a marginal eclipse depth of 33 $\pm$ 22\,ppm for TOI-431\,b in the 4.5\,µm Spitzer channel (Figure \ref{fig:bestfit}).

To ensure the robustness of our measurement, we also conducted a quick, independent analysis also using PLD. In this analysis, we first calculated the mean of each stack of 64 IRAC2 subarray frames, resulting in 689 mean frames for each of the seven AORs  (Spitzer Astronomical Observation Request). We then extracted time series from each of the central nine pixels, and constructed normalized vectors $v_{ij}$ by dividing by the sum at each timestep. To capture higher-order PLD effects, we also included a second set of vectors corresponding to $v_{ij}^2$, as well as a set of seven linear functions in time (within each AOR) to remove any long-term drifts. We then calculated the coefficients of all of these systematics vectors using linear-least-squares, using only the out-of-eclipse data. After normalizing the photometry by our optimal light curve model, we excluded 10 timesteps as outliers, because their residuals were discrepant by $>5 \sigma$. Finally, we use the average and standard error on the mean of the remaining points within, and outside of, the expected eclipse window to calculate the eclipse depth and its uncertainty. This simple analysis yields an eclipse depth of $28\pm20$~ppm, in excellent agreement with the ExoTEP result.

Assuming the planet radiates as a blackbody, we measured the dayside brightness temperature $T_{p}$ at 4.5\,µm to be $1520_{-390}^{+360}$K by inverting the following equation:
%New Tp calculated using phoenix stellar spectra integrated over ch2 bandpass instead of blackbody
\begin{equation}
    F_{p, avg} = \frac{\pi R_{p}^{2} B_{4.5\mu m}(T_{p})}{\pi R_{s}^{2} F_{s,4.5\mu m}}
\end{equation}
where $R_{p}$ and $R_{s}$ represent the radius of the planet and star respectively, $B_{4.5\mu m}(T_{p})$ is the blackbody spectrum of the planet at a temperature $T_{p}$ and wavelength $\lambda = 4.5\mu m$, and $F_{s,4.5\mu m}$ is the stellar flux integrated over the Spitzer Ch 2 bandpass, assuming a PHOENIX stellar model for TOI-431 \citep[T$_\mathrm{eff} = 4850$K, log $g = 4.6$, {[M/H]}$= 0.2$,][]{2013A&A...553A...6H}.

\section{Modeling Analysis} \label{sec:atms}
Our measured values for the eclipse depth and dayside brightness temperature of TOI-431\,b are far lower than anticipated for an atmosphereless bare-rock. A low eclipse depth may indicate high levels of heat recirculation around the planet, or a highly reflective surface. In subsection \ref{sec:grid}, we investigate the potential heat circulation on TOI-431\,b by treating it as a blackbody and investigating its heat redistribution and albedo. Subsection \ref{sec:barerock} investigates TOI-431\,b's surface composition by comparing the measured eclipse depth to emission spectra produced by bare-rock models of different mineralogical compositions. Finally, in subsection \ref{sec:scarlet}, we investigate a number of potential atmospheres on TOI-431\,b using the one-dimensional (1D) self-consistent atmosphere modeling framework SCARLET.

\subsection{Heat Redistribution around TOI-431\,b}\label{sec:grid}

\subsubsection{Methods}

The energy balance and atmospheric circulation of TOI-431\,b can be studied by investigating the Bond albedo A$_{B}$ and heat redistribution factor $f$. A$_{B}$ represents the total reflectivity index of the planet integrated over all wavelengths. We assume that the dayside flux of TOI-431 b consists entirely of light from the host star either reflected or re-emitted by the surface. This constrains the Bond albedo between 0 (no light is reflected) and 1 (all light is reflected). The heat redistribution factor $f$ represents the efficiency of the heat transport around the planet, particularly through atmospheric circulation, and can range from $\frac{1}{4}$ to $\frac{2}{3}$. A value of $\frac{1}{4}$ indicates that all incident heat is uniformly redistributed around the planet, while a value of $\frac{2}{3}$ indicates that no incident heat is redistributed, such that the planet emits the same amount of energy from its dayside as a bare-rock \citep{2008ApJS..179..484H, 48d96982-3471-3bab-9b6f-f8b21bafd6fb}. A heat redistribution factor of $\frac{1}{2}$ indicates a special case in which heat is uniformly redistributed, but only on the dayside of the planet \citep{2008ApJS..179..484H}.

One way to interpret an eclipse measurement in a single spectroscopic band is to assume the planet radiates as a blackbody to probe different combinations of A$_{B}$ and $f$ \citep[e.g.,][]{Crossfield_Malik_Hill_Kane_Foley_Polanski_Coria_Brande_Zhang_Wienke_et_al._2022}. A$_{B}$ and $f$ can used to calculate the planet's dayside equilibrium temperature:

\begin{equation}
    T_{d,eq} = T_{s} \sqrt{\frac{R_{s}}{a}}(1-A_{B})^{\frac{1}{4}}f^{\frac{1}{4}}
    \label{eq:tdeq}
\end{equation}

Using this equation, we can constrain the values of A$_{B}$ and $f$ that reproduce the measured dayside brightness temperature of TOI-431\,b.

\begin{figure}[t!]
\centering
\includegraphics[width=\linewidth]{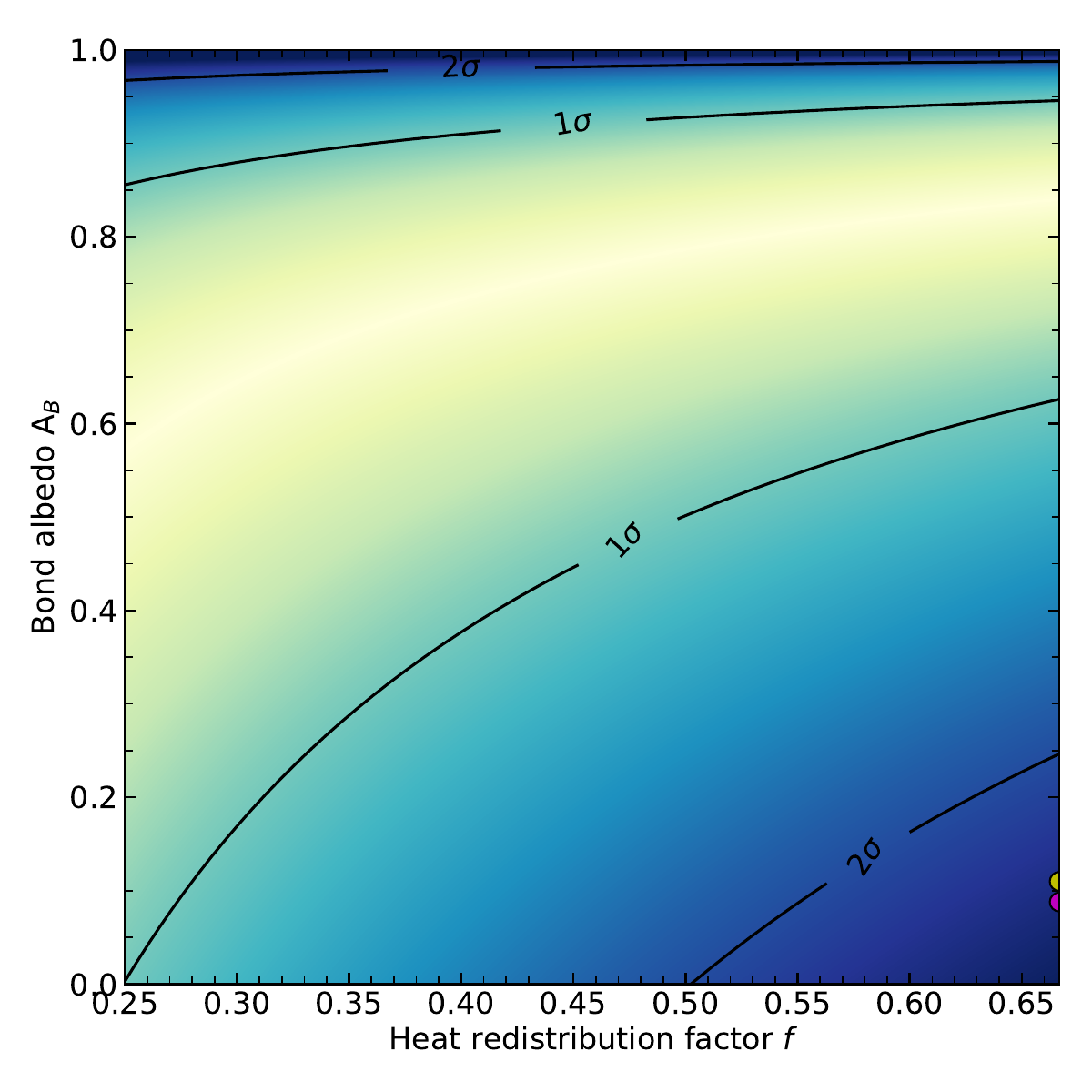}
\caption{Joint posterior distribution of the Bond albedo A$_{B}$ (vertical axis) and heat redistribution factor $f$ (horizontal axis) assuming TOI-431\,b radiates as blackbody at its 4.5\,µm brightness temperature of $1520$K. The 2D posterior is shown using colored shading, where darker regions indicate lower probabilities. The 1$\sigma$ and 2$\sigma$ regions are indicated by the black contour lines. The approximate values for the Moon and Mercury are shown in yellow and magenta at the bottom right, respectively.}

\label{fig:blackbodyAbF}
\end{figure}

\subsubsection{Results}

The 2D posterior constraining A$_{B}$ and $f$ suggests that a dark bare-rock scenario is disfavored for TOI-431\,b (Figure \ref{fig:blackbodyAbF}). Although much of the joint posterior is within $1\sigma$ of our measured brightness temperature, two sections of the posterior are disfavored. High Bond albedos (A$_{B} \lessapprox 1$) are disfavored for all values of $f$, indicating that an exotic, ultra-reflective surface composition is unlikely. More importantly, the region that gives the highest dayside temperature (A$_{B}$ = 0, $f$ = $\frac{2}{3}$) is disfavored at $>2\sigma$. Airless bodies tend to fall within this regime of minimal heat redistribution and low reflectance \citep[e.g.,][]{2018AsBio..18.1559M}. In our own solar system, airless rocky bodies are generally quite dark, with Bond albedoes of 0.11 for the Moon and 0.088 for Mercury \citep{mallama2017sphericalbolometricalbedoplanet}. 

Despite our inferences, much of the parameter space remains unconstrained. Firstly, planets with zero heat redistribution but relatively high Bond albedos are within the $1\sigma$ range, indicating that a reflective bare-rock model may be consistent with our observations. Although surface compositions with high Bond albedos are not impossible \citep[Subsection \ref{sec:barerock},][]{Hammond_Guimond_Lichtenberg_Nicholls_Fisher_Luque_Meier_Taylor_Changeat_Dang_et_al._2024} high values of A$_{B}$ are often associated with the presence of clouds \citep{Mansfield_Kite_Hu_Koll_Malik_Bean_Kempton_2019}. Secondly, the Bond albedo is largely unconstrained for models with smaller values of $f$, and thus for models with atmospheric heat recirculation. A$_{B}$ varies heavily between rocky bodies with atmospheres in our own solar system, with 0.25 for Mars and 0.76 for Venus \citep{HAUS2016178}. However, the somewhat unconstrained nature of the heat redistribution factor and the Bond albedo is not surprising, as the two parameters are strongly correlated, as shown in Equation \ref{eq:tdeq}. For example, increasing the heat redistribution factor has the same impact on T$_{d,eq}$ as decreasing A$_{B}$. 

\subsection{Bare-rock emission spectra}\label{sec:barerock}

\subsubsection{Methods}

We simulated a number of bare-rock emission spectra using the wavelength-dependent albedos of various mineralogical surfaces from \citet{Hu_Ehlmann_Seager_2012}, who used emissivity measurements of rock samples to calculate the IR spectra of each surface type \citep{clark2007usgs, cheek2009revisiting, 2001JGR...10614711W}. These models assume an airless environment and one of five surface types: ultramafic, basaltic, feldspathic, iron-oxidized, and metal-rich. These different surface compositions are indicative of different formation histories and chemical abundances \citep{Hu_Ehlmann_Seager_2012, Mansfield_Kite_Hu_Koll_Malik_Bean_Kempton_2019, Zhang_Hu_Inglis_Dai_Bean_Knutson_Lam_Goffo_Gandolfi_2024}. Magnesium-rich ultramafic rocks were common on primordial Earth and Mars, and are formed by the partial melting of rock \citep{2004E&PSL.219..173G}. Basaltic crusts associated with volcanism are abundant on the Earth and Moon today. Feldspathic crusts formed by the crystallization of magma oceans are present in the lunar highlands, while iron oxides associated with space weathering widely populate Mars. The metal-rich (pyrite) crust has no known counterpart in our solar system.

The distribution of stellar flux incident on orbiting planets is non-uniform; for an atmosphereless bare-rock, the temperature peaks at the substellar point and decreases as we approach the terminator line \citep{2008ApJS..179..484H}. Our bare-rock model accounts for the non-uniform temperature gradient on the planet's dayside by separating the hemisphere into a large number of rings, with each ring at an angle $\theta$ away from the substellar point.  The radiance of each ring is then calculated using the emissivities of the surface composition such that the sum of the reflected and thermal emission from each ring is equal to the incident net flux from the host star onto the annular section (Monaghan et al. \textit{in prep}). 

\begin{figure}[t!]
\centering
\includegraphics[width=\linewidth]{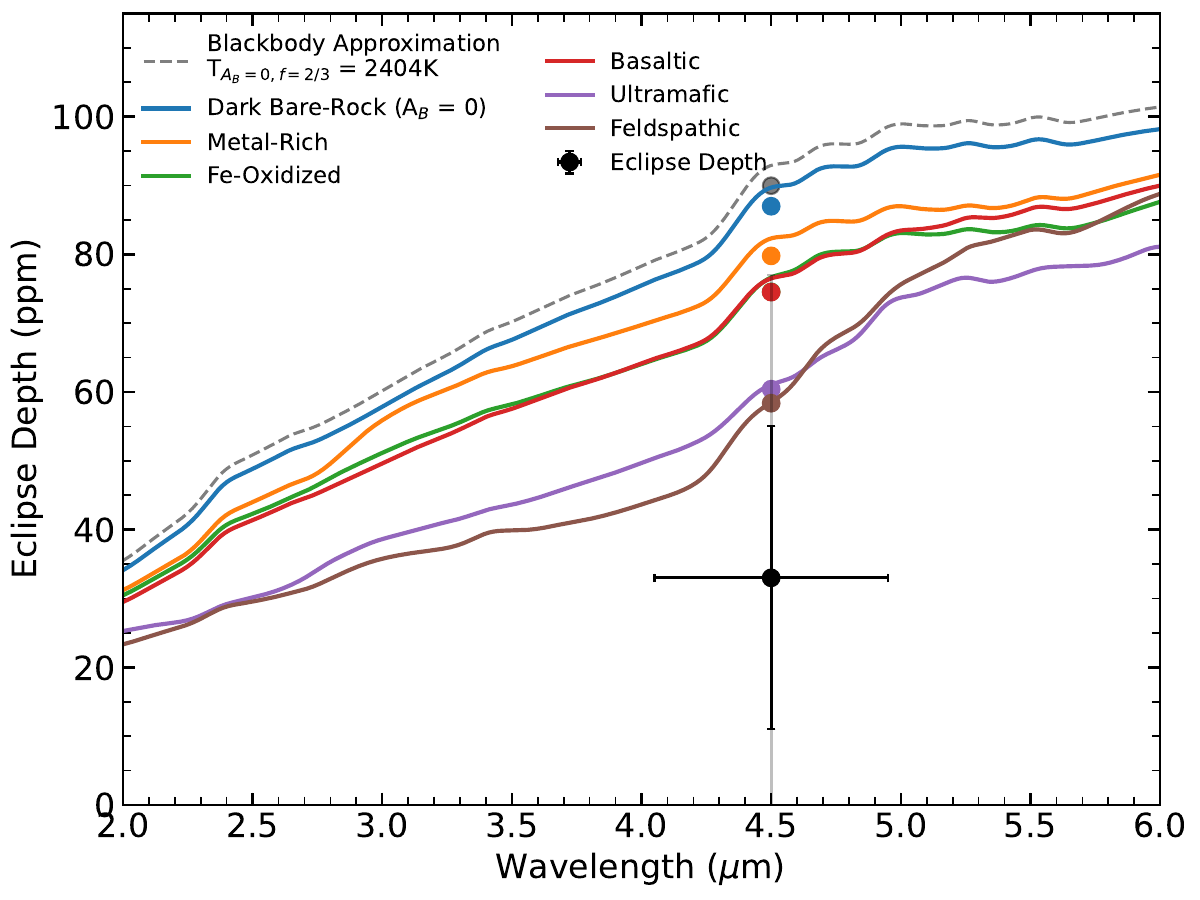}

\caption{Eclipse depth of TOI-431\,b at 4.5\,µm compared to a suite of atmosphereless emission spectra from different mineralogical surface models, assuming a PHOENIX stellar model. The black dot represents the measured eclipse depth with the associated $\pm1\sigma$ and $\pm2\sigma$ error bars shown in black and gray, respectively. The colored lines represent the emission spectra of difference surface compositions assuming a non-uniform temperature gradient on the planet's surface. Colored circles represent the models integrated over the IRAC2 bandpass between 4.05 and 4.95 \,µm. A blackbody model with a uniform dayside temperature is shown for comparison, indicating the difference between itself and the more accurate dark bare-rock model in blue.}
\label{fig:atmosphereless}
\end{figure}

\subsubsection{Results}

The non-reflective bare-rock scenarios are inconsistent to 1 $\sigma$ with the measured eclipse depth at 4.5\,µm (Figure \ref{fig:atmosphereless}). The ultramafic and feldspathic models are tenuously consistent with the data at the  $\leq 1.25\sigma$ level, while the remaining models are inconsistent to the data at the $>1.8\sigma$ level. The dark bare-rock scenario is ruled out at $2.5\sigma$. Ultramafic and feldspathic surfaces are highly reflective in the Spitzer NRS2 bandpass ($A_{B} \geq 0.4$). Our constraints on A$_{B}$ and $f$ for TOI-431\,b indicate that a high bond albedo is required for a planet with minimal heat redistribution (Figure \ref{fig:blackbodyAbF}). A feldspathic surface composition may be unsuitable for TOI-431\,b as feldspar-rich surfaces are less likely to form on Earth sized planets \citep{2012AREPS..40..113E}. However, the high temperatures on TOI-431\,b may result in a partially molten surface (see discussion below), and as such may produce a surface rich in ultramafic rock. The potential for an ultramafic surface on TOI-431\,b is of particular note, as the high albedo of ultramafic surfaces may lead to false positives for atmospheric detection, particularly around M-dwarf stars \citep{Mansfield_Kite_Hu_Koll_Malik_Bean_Kempton_2019}. 

A number of factors may affect the reflectivity of different surface compositions. Surface texture, chemical contaminants, and space weathering are all known to lower the Bond albedo of airless bodies \citep{paragas2025newspectrallibrarymodeling,    2015aste.book..597B, 2019A&A...627A..43D, lyu2024superearthlhs3844btidallylocked, Mansfield_Kite_Hu_Koll_Malik_Bean_Kempton_2019}. However, a number of factors may result in a high albedo surface without an atmospheric presence \citep[e.g.,][]{Mansfield_Kite_Hu_Koll_Malik_Bean_Kempton_2019}. Most notably, at higher surface temperatures, the partial devolatization of rock may form a highly reflective corundum surface composed of calcium and aluminum \citep{2016ApJ...828...80K}. The rate of devolatization becomes notable above 1250K, and thus hotter regions of TOI-431\,b may be affected near the substellar point \citep{Mansfield_Kite_Hu_Koll_Malik_Bean_Kempton_2019}.

Another notable complication we ignore in our bare-rock models is that the emission and reflection spectra of different mineralogical compositions is dependent on temperature \citep[e.g.,][]{2013E&PSL.371..252H, 2020E&PSL.53416089F,  2021PSJ.....2...43T, 2021Icar..35414040P}. As such, the laboratory reflectance of each composition tested may not accurately describe the conditions on TOI-431\,b. Furthermore, TOI-431\,b is potentially hot enough to melt components of its surface. Assuming zero heat recirculation and zero reflectance, the maximum dayside temperature of TOI-431\,b is $\sim$2400K as calculated by Equation \ref{eq:tdeq}. For reference, silicate rocks start melting at 850K, and all silicate rocks are completely molten by $\sim$1500K \citep[e.g.,][]{lutgens}. At our measured dayside brightness temperature of 1520K, components of TOI-431\,b's surface may be liquefied. Molten surfaces have different spectral properties from their solid counterparts, but are expected to maintain low albedos of $\leq 0.1$ \citep{Essack_Seager_Pajusalu_2020}. It is possible that a combination of molten and devolatized rock is present on the surface of TOI-431\,b, leading to a non-uniform albedo across the planet's surface.

\subsection{SCARLET Atmosphere Models}\label{sec:scarlet}

\begin{figure}[t!]
\centering
\includegraphics[width=\linewidth]{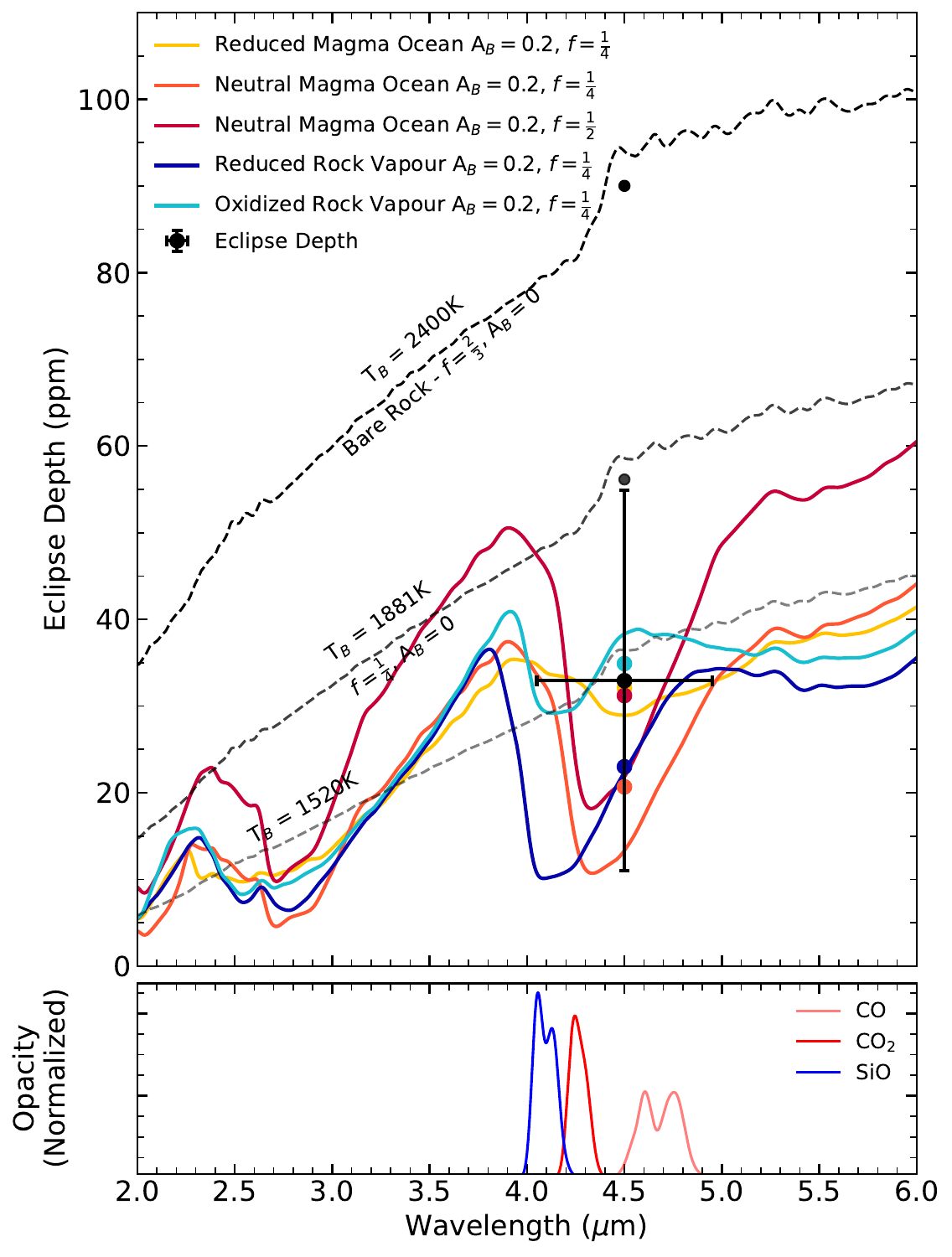}
\caption{Eclipse depth of TOI-431\,b at 4.5\,µm compared to a suite of theoretical emission spectra produced by SCARLET. The coloured lines represent the five atmospheric models simulated by SCARLET using a non-gray temperature profile and a well-mixed composition. The dashed gray lines show the blackbody eclipse spectra corresponding to different combinations of $f$ and A$_{B}$ at uniform dayside temperatures. The black dot represents the measured eclipse depth, with the associated $\pm1\sigma$ error bars. Coloured circles represent the models integrated over the IRAC2 bandpass between 4.05 and 4.95 \,µm. The exact compositions of the atmospheric models are shown in Table \ref{tab:models}. Normalized opacity bands from dominating molecules in the models (CO, CO$_{2}$, and SiO) are shown in the bottom panel.}
\label{fig:atmospheremodels}
\end{figure}

\subsubsection{Methods}

We used the SCARLET atmosphere modeling frame- work to generate a suite of 1D self-consistent atmosphere models of TOI-431\,b and compare the model emission spectra to the Spitzer eclipse measurements \citep[e.g.,][]{2012ApJ...753..100B, 2013ApJ...778..153B, benneke2015strictupperlimitscarbontooxygen, 2019ApJ...887L..14B, 2019NatAs...3..813B, benneke2024jwstrevealsch4co2, 2021AJ....162...73P, pelletier2024criresespressorevealatmosphere,Roy_Benneke_Piaulet_Crossfield_Kreidberg_Dragomir_Deming_Werner_Parmentier_Christiansen_et_al._2022, Roy_2023, Piaulet_Ghorayeb_2024, Bazinet_2024}. We used the framework to generate non-gray models of TOI-431\,b assuming a well-mixed composition. The wavelength-dependent models calculate the chemical composition of the planet by solving for hydrostatic equilibrium and radiative transfer iteratively to produce a theoretical temperature-pressure profile of the planet. Once the model converges to a stable temperature profile and chemistry, the associated emission spectrum from the secondary eclipse is computed. The stellar spectrum is accounted for in the model using the same PHOENIX stellar emission spectrum described earlier.

Our suite of models studies a variety of atmospheric compositions that may be generated on TOI-431\,b. We test two atmospheres produced by volatile magma oceans, and two metal-rich rock vapor atmospheres. The exact compositions of the models simulated are shown in Table \ref{tab:models}.

\begin{table}[t]
\centering
\footnotesize
\begin{tabular}{ c | c c c}
\hline
Model   & pCloud &  Composition  & Ref \\
& (bar) & (abundance) & \\
\hline
\hline
Reduced Magma & & H$_{2}$: 42\%    CO: 42\% & \\
Ocean & 1 & N$_{2}$: 15\%    H$_{2}$O: 1\% & 1 \\
\\
\hline
Neutral Magma  & & CO: 35\%    CO$_{2}$: 30\%& \\ 
Ocean& 1 & N$_{2}$: 30\%    H$_{2}$O: 4\% & 1\\ 
& &  H$_{2}$: 1\% & \\ 
\hline
 & & SiO: 46.4\%    Mg: 28.2\% & \\
Reduced Rock & & Fe: 25.1\%    Si: 0.1\% & \\
vapor & $10^{-4}$ & O: 0.1\%& 2\\
 & & Trace SiO$_{2}$, MgO, FeO & 
\\
\hline
Oxidized Rock & & O: 74.6\%    O$_{2}$: 25.1\%& \\
vapor &$10^{-4}$ & Fe: 0.01\%    FeO: 0.01\% & 2\\
 & & SiO: 0.01\%    Mg: 0.01\% & \\

\hline
\hline
\end{tabular}
\caption{\label{tab:models} Summary of atmospheric models shown in Figure \ref{fig:atmospheremodels} as computed by SCARLET, alongside the associated source for each. 'Trace' implies an abundance of 0.001\% \newline
1: \citet{Maurice_Dasgupta_Hassanzadeh_2024} \newline
2: \citet{Seidler_Sossi_Grimm_2024}}
\end{table}

The chemical abundances of the magma outgassed atmospheres were approximated from Figure 3 of \citet{Maurice_Dasgupta_Hassanzadeh_2024}, which represents the compositions produced by lava oceans of different oxidation states. Although such compositions represent a primordial magma ocean during the early lifetime of an exoplanet, they may approximate the chemistry associated with a permanent magma ocean rich with volatiles. We model the atmospheres produced on a 3M$_{\oplus}$ planet by a heavily reduced magma ($\Delta$IW = -5) and by a neutral magma (Earth-like, $\Delta$IW = 0), assuming both oceans are 0.5 times the planetary silicate mass.

We test the neutral magma ocean model at two levels of heat redistribution: one with full recirculation to the nightside of the planet ($f = \frac{1}{4}$), and one with uniform recirculation on the dayside only ($f = \frac{1}{2}$). The inclusion of a model without nightside atmospheric circulation is done to simulate a planet where the atmosphere condensates or solidifies on the nightside due to the colder temperatures, and to showcase that a thick, entirely redistributive atmosphere may not be required to produce the observed eclipse depth of TOI-431\,b \citep[e.g.,][]{2018A&A...617A.110P, Ehrenreich_2020}.

The chemical abundances of the rock vapor atmospheres were approximated from Figure 6 of \citet{Seidler_Sossi_Grimm_2024}, which represents the mineral atmospheres produced by a silicate-rich magma using a bulk silicate Earth-like mantle composition model. To specify the abundances of the well-mixed composition type, we choose to use the mixing ratios simulated at an irradiation temperature of 2000K and $10^{-4}$ bars of pressure for a highly oxidized environment ($\Delta$IW = 4) and a highly reduced state ($\Delta$IW = -4).

\subsubsection{Results}

The model emission spectra for the rock-vapor and magma-ocean atmospheres agree well with our measured eclipse depth within $1\sigma$ (Figure \ref{fig:atmospheremodels}). Most notably, the neutral magma ocean with uniform dayside redistribution ($f = \frac{1}{2}$) is shown to be consistent due to the large absorption band at $\sim4.5\mu$m, despite the lack of atmospheric recirculation to the nightside of the planet.  The blackbody model at $f = \frac{1}{4}$, A$_{B} = 0$ is shown to be tenuously consistent at $\sim 1 \sigma$.

There are a number of potential caveats to our SCARLET models. The presence of clouds, winds, and photochemistry may all affect the temperature-pressure profile and composition of TOI-431\,b in complex ways. A number of cloud species have condensation temperatures around the dayside brightness temperature of TOI-431\,b, and as such could be present on the planet \citep{2002ApJ...577..974L, Wakeford_2016}. We further note that the chosen chemical abundances simulated in our well-mixed compositions are only approximations. \citet{Maurice_Dasgupta_Hassanzadeh_2024} simulates the volatile-rich atmospheres at a much higher pressure than those expected for TOI-431\,b, while the mixing ratios simulated by \citet{Seidler_Sossi_Grimm_2024} are sensitive to a range of potential atmospheric pressures. These models also generate different temperature-pressure profiles than the ones generated by SCARLET. The rock vapor profiles generated by \citet{Seidler_Sossi_Grimm_2024} in particular are inverted due to the presence of other optical absorbers in the simulated atmospheres. These potential caveats are unlikely to affect our results due to the large uncertainties in our measured eclipse depth. The models should remain consistent under small modifications to the chemical abundances or temperature profile.

%I don't like "captured". We didn't capture the individual eclipses. Rewrite this whole thing in more detail. Write an entire subsection on the absence of the eclipse that is similar to the response to the referee. and instead of "captured" say "our observations to cover the time interval during which the planet is eclipsed by the host star."

\section{Discussion} \label{sec:dis}
Our combined atmospheric analysis of TOI-431\,b disfavors a low-albedo bare-rock composition at 2.5$\sigma$, hinting towards either an atmospheric presence or a highly reflective bare-rock surface. However, probing the composition of a planetary atmosphere from an eclipse depth in a single wavelength regime does not place stringent constraints on the planet's composition. We discuss the results of our data reduction and analysis below. In subsection \ref{sec:interpreting}, we discuss the significance of TOI-431 b's low eclipse depth. We then consider the potential atmospheric scenarios that may produce a shallow eclipse depth in subsection \ref{sec:potentialatms}. Finally, in subsection \ref{sec:followups}, we discuss the potential for follow-up observations of TOI-431\,b, and what they may reveal regarding the nature of the planet.

\subsection{Interpreting the Absence of a Deep Eclipse} \label{sec:interpreting}

The occultation depth of 33$\pm$22\,ppm measured for TOI-431\,b is not a significant detection ($1.5\sigma$). However, the absence of a deep eclipse disfavors the bare-rock scenario, which suggests an eclipse depth of $\sim$ 90\,ppm in the IRAC2 bandpass of Spitzer. If the eclipse depth of the planet were this deep, our 22ppm precision would have resulted in a detection at $>4\sigma$. Thus, we have the precision to strongly detect the eclipse of a bare-rock composition.

Another factor to consider is the possibility that the occultation events were not observed during each Spitzer visit due to a highly eccentric orbit. An eccentric orbit for TOI-431 b would be unexpected due to its proximity to the host star. Closely orbiting planets like TOI-431 b generally have negligible eccentricities due to tidal circularization \citep[e.g.,][]{Rasio:1996nm, Marcyyyyy, Rodr_guez_2011}. We estimate the circularization timescale of TOI-431 b using equation 2 provided by \citet{2008ApJ...678.1396J}. Depending on TOI-431 b’s love number $k_{2}$ and quality factor $Q$, the circularization timescale is $\leq$ 6 Myr using Jupiter’s values of $k_{2}$ and $Q$, which likely underestimate the tidal forces exerted on TOI-431 b \citep{2008ApJ...678.1396J, Lainey_2016}. Comparing this timescale to the estimated age of the TOI-431 system (1.9 Gyr) indicates that the orbit of TOI-431 b is highly unlikely to be eccentric, in agreement with the result based on the radial-velocity data \citep{Osborn_2021}.

To further ensure the occultation events were observed in each Spitzer visit, we estimated the maximum phase offset of TOI-431 b's eclipse assuming an eccentric orbit. In the discovery of TOI-431 b, \citet{Osborn_2021} measure a 95\% confidence interval from 0 to 0.28 for the planet's eccentricity. The phase of a secondary eclipse event can be calculated with:

\begin{equation}
    phase = \frac{1}{2} + \frac{2}{\pi}e\,cos(\omega)
\end{equation}

\noindent where $e$ is the eccentricity and $\omega$ the argument of periastron \citep{winn2014transitsoccultations}. With this equation, we measure a maximum timing offset from $\frac{1}{2}$ phase of 125 minutes, assuming $e = 0.28$ and $\omega = 0$. With each Spitzer visit centered on phase $\frac{1}{2}$ and lasting 5.24 hours, the occultation events would be observed at least partially for all observations.

\subsection{Potential Atmospheric Scenarios} \label{sec:potentialatms}
A significant atmosphere on TOI-431\,b would be difficult to maintain for a long lifetime. XUV irradiation from its host star is theorized to drive atmospheric photoevaporative escape at rates of $10^{6} - 10^{8}$ kg/s, far higher than the CO$_{2}$ outgassing rate of Earth \citep{King_Corrales_fern_Wheatley_Malsky_Osborn_Armstrong_2024, 2019Natur574yeah, jiang2025estimatemassescapingrates}. Although a rapidly outgassing magma ocean may exceed the mass loss rate, the outgassing rate on such planets is dependent on a number of factors, and is difficult to constrain \citep[e.g.,][]{Dorn_2018, 2020A&A...643A..44S, 2023A&A...675A.122B}.

The 'cosmic shoreline' theory proposes that a planet's ability to retain an atmosphere is based on the predicted escape velocity and XUV irradiation \citep{Zahnle_2017}. TOI-431\,b is expected to receive $\sim 7 \times 10^{4} - 9 \times 10^{4}$ erg/s cm$^{2}$ of XUV irradiation, placing it well above the shoreline \citep{King_Corrales_fern_Wheatley_Malsky_Osborn_Armstrong_2024, jiang2025estimatemassescapingrates}. However, 55 Cnc e is similarly placed well above the cosmic shoreline, despite evidence to suggest the planet hosts a volatile-rich atmosphere \citep{Hu_Bello-Arufe_Zhang_Paragas_Zilinskas_Van_Buchem_Bess_Patel_Ito_Damiano_et_al._2024, Patel_Brandeker}. Furthermore, TOI-431\,b is also exposed to high levels of XUV irradiation, and suffers atmosphere loss at a rate of $10^{7}$ kg/s \citep{heng2023transientoutgassedatmosphere55}. As such, the planet may instead host a transient atmosphere, which could cause the apparent variability seen in phase curve observations \citep{heng2023transientoutgassedatmosphere55, Patel_Brandeker}. Further evidence of flux variability of TOI-431 b in the optical and infrared regimes would be necessary to determine the presence of transient outgassing.

\subsection{Follow-up Observations} \label{sec:followups}

Follow-up studies of TOI-431\,b's emission spectrum are necessary to confirm the low occultation depth and investigate an atmospheric presence. A number of spectral features may be used to analyze the volatile content in spectroscopic analysis. A volatile-rich atmosphere with CO/CO$_{2}$ can be identified by its large absorption between 4-5\,µm, while a mineral-rich atmosphere may be identified using the SiO feature visible at 4.2\,µm (Figure \ref{fig:atmospheremodels}). Spectral features may be identified at longer wavelengths, including MgO at 6\,µm, SiO at 9\,µm, and CO$_{2}$ at 15\,µm \citep[e.g.,][]{Seidler_Sossi_Grimm_2024, Rieke_2015}. 

Identifying molecular signatures on TOI-431\,b requires the use of spectroscopy, which is readily achieved using a number of instruments on the \textit{JWST}. Comparing the observed eclipse depth within these wavelength regimes may constrain the bulk volatile content and oxidation of TOI-431\,b's atmosphere, and may probe the temperature-pressure profile of the planet itself. Further information regarding the planet's energy budget, heat redistribution, and two-dimensional structure may be inferred through phase curve observations, which could also be used to analyze the redistribution of heat around the planet.

If an atmosphere is present on TOI-431\,b, constraining its composition would significantly further our under-
standing of the potential atmospheres a planet can retain under high irradiation. The models simulated above were chosen due to their similarities to atmospheres proposed for other rocky exoplanets. Recent observations of the emission spectra of 55 Cnc e may indicate the presence of a similarly volatile atmosphere composed of CO or CO$_{2}$ sustained by the outgassing of a magma ocean \citep{Hu_Bello-Arufe_Zhang_Paragas_Zilinskas_Van_Buchem_Bess_Patel_Ito_Damiano_et_al._2024, Patel_Brandeker}. Furthermore, rock vapor atmospheres have previously been proposed for other hot USPs, including K2-141 b and 55 Cnc e \citep{Zieba_Zilinskas_Kreidberg_Nguyen_Miguel_Cowan_Pierrehumbert_Carone_Dang_Hammond_et_al._2022, rasmussen2023nondetectionironhighresolutionemission}. The prevalence and nature of rock vapor atmospheres will be further studied as part of GO 4818 (PI: Mansfield).

%%Combine information here with the paragraph(s) in figure 
\section{TOI-431\,b in Context} \label{sec:context}
TOI-431\,b joins a growing list of rocky planets ($\leq$2R$_{\oplus}$) with measured infrared thermal emissions. The notably low dayside emission interrupts a trend first noted by \citet{Crossfield_Malik_Hill_Kane_Foley_Polanski_Coria_Brande_Zhang_Wienke_et_al._2022} of smaller planets having temperatures consistent with bare-rock models \citep[e.g, ][]{Greene_Bell_Ducrot_Dyrek_Lagage_Fortney_2023, Zieba_Kreidberg, Zhang_Hu_Inglis_Dai_Bean_Knutson_Lam_Goffo_Gandolfi_2024, Xue_Bean_Zhang_Mahajan_Ih_Eastman_Lunine_Mansfield_Coy_Kempton_et_al._2024, Mansfield_Xue_Zhang_Mahajan_Ih_Koll_Bean_Coy_Eastman_Kempton_et_al._2024, wachiraphan2024thermalemissionspectrumnearby, luque2024darkbarerocktoi1685, Ducrot_2024}. TOI-431\,b is now the fourth rocky exoplanet with thermal emissions showing some evidence for heat redistribution, alongside 55 Cnc e, K2-141 b, and LHS 1478 b \citep{Hu_Bello-Arufe_Zhang_Paragas_Zilinskas_Van_Buchem_Bess_Patel_Ito_Damiano_et_al._2024, Zieba_Zilinskas_Kreidberg_Nguyen_Miguel_Cowan_Pierrehumbert_Carone_Dang_Hammond_et_al._2022, august2024hotrockssurveyi}. While LHS 1478 b orbits a relatively cool M dwarf, the other three planets are subject to the highest levels of irradiation from their host stars when compared to other planets with thermal emission observations. This may indicate that highly irradiated planets outgas secondary atmospheres more efficiently. Furthermore, these three planets are known to orbit relatively hot stars: 55 Cnc e orbits a G star, while K2-141 b and TOI-431\,b both orbit K stars. The potential for atmospheric retention may therefore be impacted by the host star itself. Many planets observed in the infrared regime through secondary eclipses orbit M stars. M dwarfs are generally active; stellar winds, coronal mass ejections, and XUV irradiation may contribute to the atmospheric erosion of orbiting bodies \citep{Mignon_2023}. However, the thermal emission measured from the super-Earth LHS 1478 b indicates a brightness temperature inconsistent with the bare-rock hypothesis, despite the planet orbiting an M-dwarf star \citep{august2024hotrockssurveyi}. It should be noted that the results from LHS 1478 b are based on a single eclipse visit. 

Further study is required to fully understand the presence and composition of rocky planet atmospheres. The trends discussed above are tenuous at best, largely due to the lack of data points necessary for reliable conclusions to be made. A variety of factors may play a role in retaining the atmospheres of rocky bodies, including stellar type and planetary radius. Future and ongoing observations by the \textit{JWST} will expand the scope of research into rocky planet atmospheres through secondary eclipse and phase curve observations. GO 4818 (PI: Mansfield) will investigate the formation and properties of silicate atmospheres using MIRI/LRS eclipse observations of the potential lava worlds GJ 9827 b, HD 20329 b, TOI-1416 b, TOI-500 b, TOI-1442 b, TOI-561 b, TOI-1075 b, TOI-1807 b, HD 3167 b, and TOI-431\,b. The atmospheres of rocky M dwarf exoplanets will also be investigated as part of the proposed 500 hours \textit{JWST} DDT program, which has so far been confirmed to observe LTT 1445A c and GJ 3929 b \citep{Redfield_Batalha_Benneke_Biller_Espinoza_France_Konopacky_Kreidberg_Rauscher_Sing_2024}. Among these larger surveys, the \textit{JWST} will also observe TOI-2445A b (GO 3784), TOI-4481 b (GO 4931), and LP-781-18 d (GO 6457).

\section{Conclusion} \label{sec:con}
TOI-431\,b is among the first terrestrial planets with infrared thermal emission observations to show a notably low eclipse depth. Our modeling of the planet's composition disfavors a low-albedo bare-rock scenario, and instead suggests the presence of an atmosphere potentially fueled by the outgassing of a magma ocean. Alternatively, the planet's low eclipse depth may be produced by a bare-rock whose surface is highly reflective. Constraining the exact composition of the planet, however, requires further study with modern instrumentation and it remains to be understood how an atmospheric presence on TOI-431\,b could be maintained under the intense XUV radiation from its host star. Future observations of TOI-431\,b offer the opportunity to further investigate the nature of this planet and further probe its surface and atmospheric composition. We encourage further research into the infrared emission spectra of rocky exoplanets to understand their compositions and potential for atmospheric retention.

\acknowledgements
We wish to thank the anonymous referee for their report, which greatly enhanced the quality of the manuscript. This work is based on observations made with the Spitzer Space Telescope, which was operated by the Jet Propulsion Laboratory, California Institute of Technology, under a contract with NASA. C.M. and P.-A. R. acknowledge financial support from the University of Montreal, and C.M. further acknowledges financial support from Jean-Marc Lauzon. B.B., P.-A.R., and C.M. acknowledge financial support from the Natural Sciences and Engineering Research Council (NSERC) of Canada. C.P.-G acknowledges support from the NSERC Vanier scholarship, and the Trottier Family Foundation. C.P.-G also acknowledges support from the E. Margaret Burbidge Prize Postdoctoral Fellowship from the Brinson Foundation. This work was made with the support of the Institut Trottier de Recherche sur les Exoplanetes (iREx).

\bibliography{main}

%% This command is needed to show the entire author+affilation list when
%% the collaboration and author truncation commands are used.  It has to
%% go at the end of the manuscript.
%\allauthors

%% Include this line if you are using the \added, \replaced, \deleted
%% commands to see a summary list of all changes at the end of the article.
%\listofchanges

\end{document}